\documentstyle[12pt]{article}
\advance \textheight by \topskip
\begin{document}
\begin{titlepage}
\hbox to \hsize{\hfil IHEP 96--29}
\hbox to \hsize{\hfil hep-ph/9604262}
\hbox to \hsize{\hfil April, 1996}
\vfill
\large \bf
\begin{center}
Would unitarity slow down the rise of $F_2(x,Q^2)$ at $x\rightarrow 0$?
\end{center}
\vskip 1cm
\normalsize
\begin{center}
{\bf S. M. Troshin and N. E. Tyurin}\\[1ex]
{\small  \it Institute for High Energy Physics,\\
Protvino, Moscow Region, 142284 Russia}
\end{center}
\vskip 1.5cm
\begin{abstract}
On the grounds of the $U$--matrix form of $s$--channel unitarization, we
consider constraints unitarity provides for the total cross--section
of the virtual photon--induced reactions and discuss the preasymptotic
nature of hadron and photon scatterings at $\sqrt{s}\leq 0.5$ TeV.
\end{abstract}
\vfill
\end{titlepage}
\section*{Introduction}
Recent HERA data \cite{h1,zeus}  have clearly demonstrated rising
 behavior
of the structure function $F_2(x,Q^2)$ at small $x$ and led to many
interesting discussions \cite{land}
on the Pomeron structure
in soft and hard interactions.
These data can be interpreted as the dependence of
 the virtual photon--proton
total
cross--section
 $\sigma_{tot}^{\gamma^*p}(W)$
 on center of mass energy $W$.
The observed rise
of $\sigma_{tot}^{\gamma^*p}(W)$
 is consistent with a linear dependence
on $W$ and has been treated somewhere as a manifestation of
 hard BFKL Pomeron \cite{lipa}.
This linear rise of $\sigma_{tot}^{\gamma^*p}(W)$
 has been considered to a somewhat extent as a surprising fact
on the grounds of our knowledge of the
 energy dependence of total cross--sections in hadronic
 interactions.

Indeed, the above comparison between photon--induced
and hadron--induced interactions is quite legitimate since
the photon is demonstrating its hadronlike nature for a long time.
It has been
accounted by the Vector Dominance  Model (VDM) \cite{vdm}.  The apparent
contradiction between the hadron and virtual photon total cross--section
behaviors has no fundamental meaning in the preasymptotic energy region where
the Froissart--Martin bound does not restrict the particular form of hadronic
 cross--sections.  Indeed, it has been shown that the total hadronic
cross--sections rise is, in fact, consistent with the linear dependence on
$\sqrt{s}$ \cite{preas} up to $\sqrt{s}\leq 0.5$ TeV and thus there is no
qualitative disagreement with the trends observed in the
$\sigma_{tot}^{\gamma^*p}(W)$ at HERA energies ($W\leq 0.3$ TeV).

 At higher energies,
 unitarity leads to deviation from the linear rise with
$\sqrt{s}$ in  hadronic total cross--sections.
  It converts a powerlike
preasymptotic increase into asymptotic $\log^2s$ rise.

However,
it could not be the case for $\sigma_{tot}^{\gamma^*p}(W)$ since the
asymptotic theorems can not be directly applied for the virtual photon
scattering. In this context, a practice to treat the virtual photon
 similar to the on--shell hadron
can not be considered as a generally valid because of the off--shell
effects.

 The
 problem was addressed in \cite{petrov} on the basis of
unitarity for off--shell scattering starting from the eikonal
representation for the scattering amplitude. It was argued that
the observed rise of $F_2(x,Q^2)$ at small $x$ or
 $\sigma_{tot}^{\gamma^*p}(W)$ rise at large $W$ values can be considered
as a  true asymptotic behavior and extension of the eikonal representation
for off-shell particles does not provide limitations for the structure
function $F_2$ at $x\rightarrow 0$.

In this paper we treat similar problems on the basis of the
  $U$--matrix
approach to the scattering amplitude unitarization. We obtain the unitary
representation for the case of off--shell particles and discuss
assumptions when unitarity does provide limitations for the amplitude of
off--shell particle scattering. We also consider specific model
parameterization for the $U$--matrix obtained on the basis of chiral quark
model in \cite{csn} and discuss corresponding results for the structure
 function $F_2(x,Q^2)$ at small $x$ values.

\section{Off--shell scattering amplitude in the $U$--matrix
approach}
In the process of photon--hadron interactions there is a significant
probability for photon coupling directly to vector mesons, i.e. photon
fluctuates into intermediate vector meson states which afterwards interact
with a hadron. It can be expressed as the field current
identity \cite{kroll}:
\begin{equation}
J_\mu=\sum_V\frac{e}{f_V}m_V^2V_\mu,\label{id}
\end{equation}
where $J_\mu$ is the electromagnetic current and $V_\mu$ --- vector
meson fields ($V=\rho,\,\omega,\,\varphi,\,J/\psi$), $f_V^{-1}$
is the $\gamma$-$V$ coupling constant and $e=\sqrt{4\pi\alpha_{em}}$.
Owing to this identity the amplitude of virtual-photon Compton
scattering can be represented in the form:
\begin{equation}
F_{\gamma\gamma}=
\sum_{V,V'}\left(\frac{e}{f_V}\right)F_{VV'}
\left(\frac{e}{f_{V'}}\right),\label{vdm}
\end{equation}
where $F_{VV'}$ is the  scattering amplitude of the process
$VN\rightarrow V'N$.
We use diagonal approximation
\[
F_{VV'}=
F_{VV}
\delta_{VV'}
\]
since we expect that diffraction dissociation amplitudes can be neglected
compare to elastic scattering amplitudes \cite{ashd}.
The amplitudes $F_{VV'}$ in Eq. (\ref{vdm}) describe scattering of off--shell
vector mesons. We consider for a while a single vector meson field
and denote $F^{**}(s,t,Q^2)$, $F^{*}(s,t,Q^2)$ and $F(s,t)$
the amplitudes when both mesons (initial and final) are off mass shell, only
initial meson is off mass shell and both mesons are on mass shell,
correspondingly.  The virtualities of initial
and final vector mesons were chosen equal $Q^2$ since we
will need further the amplitude of the forward virtual Compton scattering.

Our aim now is to consider unitarity constraints for the amplitudes
$F^*$ and $F^{**}$. We use the $U$--matrix form of unitary
representation for the scattering amplitude. It grounds
on the relativistic generalization of the Heitler equation in the
theory of radiation damping \cite{heit}. The $U$--matrix equation has been
derived in the relativistic theory in \cite{logn} in the framework
of the single--time formalism in QFT. It provides for the scattering
 amplitude
of scalar on--shell particles simple algebraic form in impact
parameter representation:  \begin{equation} F(s,b)=\frac{U(s,b)}{1-iU(s,b)}
\label{um} \end{equation} where $U(s,b)$ is the generalized reaction matrix.
It is considered as an input dynamical quantity similar, e.g. to the eikonal
function.  The inelastic overlap function is connected with $U(s,b)$ by the
relation \begin{equation}
\eta(s,b)=Im U(s,b)|1-iU(s,b)|^{-2}.
\end{equation}
Eq. (\ref{um}) ensures $s$--channel unitarity provided that
$Im U(s,b)\geq 0$.

 Eq. (\ref{um}) has completely different analytical structure
as compared to eikonal form, in particular, it does not generate
essential singularity in the complex $s$--plane at infinity while
the eikonal representation does.

It is to be noted that the solution of unitarity in potential
scattering for the case of off--shell particles in the
$K$--matrix form (the $U$--matrix is its relativistic analog)
  was given for the first time in \cite{lavl}.

Eq. (\ref{um}) was obtained  on the basis of the
relation between the matrix element of the radiation operator
\[
R(x_1,x_2;y_1,y_2)=-
\langle 0|\frac{\delta^4S}{
\delta \varphi_1^*(x_1)
\delta \varphi_2^*(x_2)
\delta \varphi_1(y_1)
\delta \varphi_2(y_2)}S^+|0\rangle
\]
and the function $U(x_1,x_2;y_1,y_2)$ which parameterizes the evolution
operator \cite{logn}. In the momentum space the relation has the form:
\begin{eqnarray}
R(p_1,p_2;q_1,q_2) & = & U(p_1,p_2;q_1,q_2)+ \nonumber\\
 & & \frac{1}{2(2\pi)^3}\int dk_1dk_2U(p_1,p_2;k_1,k_2)D^-(k_1)D^-(k_2)
R(k_1,k_2;q_1,q_2) \label{r}
\end{eqnarray}
where
 \[ D^-(k)=2\pi i \theta (k^0)\delta(k^2-m^2).  \] In Eq. (\ref{r}) only
 momenta of intermediate particles $k_1$ and $k_2$  always lie on the mass
shell while the momenta of external particles can be shifted from the mass
shell. Following Ref. \cite{logn} one can easily obtain equations for the
amplitudes $F^{**}$ and $F^*$. These equations have the same
structure as the equation for the on--shell amplitude $F$
but relate the different amplitudes.
 In the impact
parameter representation ($s\gg 4m^2$) they can be written as follows
\begin{eqnarray}
F^{**}(s,b,Q^2) & = & U^{**}(s,b,Q^2)+iU^{*}(s,b,Q^2)F^{*}(s,b,Q^2)\nonumber\\
F^{*}(s,b,Q^2) & = & U^{*}(s,b,Q^2)+iU^{*}(s,b,Q^2)F^{}(s,b).\label{es}
\end{eqnarray}
The solutions  are
\begin{eqnarray}
F^*(s,b,Q^2) & = & \frac{U^*(s,b,Q^2)}{1-iU(s,b)},\label{*}\\
F^{**}(s,b,Q^2) & = & \frac{U^{**}(s,b,Q^2)}{1-iU(s,b)}+
i\frac{[U^{*}(s,b,Q^2)]^2-U^{**}(s,b,Q^2)U(s,b)
}{1-iU(s,b)}.\label{**}
\end{eqnarray}
 Eq. (\ref{*}) is quite similar to Eq. (\ref{um}).
 The
appearance of the second term in Eq. (\ref{**}) reflects the  role of
off--shell effects. If this term is different from zero then we would arrive
 to the conclusions made in \cite{petrov} on the basis of the generalization of
eikonal representation for the off--shell particles, i. e. unitarity does not
lead to the  constraint $|F^{**}(s,b,Q^2)|\leq 1$ and consequently
 a power--like asymptotic rise of the cross--sections does not contradict
to unitarity.

However, there is another possibility, when
\begin{equation}
[U^{*}(s,b,Q^2)]^2-U^{**}(s,b,Q^2)U(s,b)=0.\label{zr}
\end{equation}
We consider the latter  in some details.
Eq. (\ref{zr}) will be fulfilled identically if the following
factorization occurs:
\begin{eqnarray}
U^{**}(s,b,Q^2) & = & \omega(s,b,Q^2)U(s,b)\omega(s,b,Q^2)\nonumber\\
U^{*}(s,b,Q^2) & = & \omega(s,b,Q^2)U(s,b).\label{fct}
\end{eqnarray}
Such factorization is valid, e. g. in the Regge model with factorizable
residues and the $Q^2$--independent trajectory.

Eq. (\ref{zr}) is also valid in VDM if
 the fluctuation length $d_{fluct}$ is significantly longer
than the characteristic size of strong interactions, i.e.
\begin{equation}
d_{fluct}\simeq\frac{2\nu}{Q^2+m_V^2}\gg 1 fm. \label{apl}
\end{equation}
The fluctuation length is proportional to the inverse of minimum energy
needed to put the vector meson on its mass shell. If Eq. (\ref{apl})
is valid
we can treat interacting vector meson as an on--shell hadron. Then
Eq. (\ref{fct}) is also valid and the function $\omega$ is proportional
to the vector meson propagator, i.e.
\begin{equation}
\omega (Q^2)=D_V(Q^2)=\frac{m_V^2}{Q^2+m^2_V}.\label{prp}
\end{equation}
Thus, we will have for $F^*$ and $F^{**}$ the following representations
\begin{eqnarray}
F^{*}(s,b,Q^2) & = & \frac{U^{*}(s,b,Q^2)}{1-iU(s,b)}=
D_V(Q^2)\frac{U(s,b)}{1-iU(s,b)}
\label{vrq}\\
F^{**}(s,b,Q^2) & = & \frac{U^{**}(s,b,Q^2)}{1-iU(s,b)}=
D^2_V(Q^2)\frac{U(s,b)}{1-iU(s,b)}
\label{vr}
\end{eqnarray}
and unitarity will provide
\begin{eqnarray}
|F^*(s,b,Q^2)| & \leq & D_V(Q^2),\nonumber\\
|F^{**}(s,b,Q^2)| & \leq & D_V^2(Q^2).\label{bnd}
\end{eqnarray}
Thus, the unitarity does restrict the amplitudes $F^{**}$ and $F^*$ in
the kinematical region where Eq. (\ref{apl}) is valid, i.e. at
not too high $Q^2$--values.
Therefore in the case of $\gamma^* p$--interactions there is  a range of
$Q^2$ where the general solution of unitarity Eqs. (\ref{*}) and
(\ref{**}) are reduced to a simpler form. It seems relevant and important
for the behavior of the structure function $F_2$ at small values of $x$.

To discuss this issue
 we consider a specific model for
hadron scattering based on the ideas of chiral quark models \cite{csn}.  In
the model valence quarks located in the central part of hadron are supposed to
scatter in a quasi-independent way by the mean field generated by
the virtual massive quarks and by the selfconsistent field of valence quarks
themselves.  In accordance with the quasi-independence of valence quarks the
 $U$--matrix is represented as the product:  \begin{equation} U(s,b)\,=\,
\prod^N_{q=1}\,\langle f_q(s,b) \rangle\label{fact}  \end{equation} in the
impact parameter representation.  Factors $\langle f_q(s,b)\rangle$ correspond
to the averaged individual quark scattering amplitude in the mean field.  of a
 Eq. (\ref{fact}) implies that all valence quarks are
scattered in the mean field simultaneously.  Such factorization  could be
considered as an effective implementation of constituent quarks' confinement.
This mechanism resembles the Landshoff mechanism of quark--quark independent
  scattering \cite{lands}.  However, in our case we refer not to pair
 interaction of valence quarks from the two colliding hadrons, but
rather to Hartree--Fock approximation for the scattering of valence quark in
the mean field.

The $b$--dependence of the function $\langle f_q \rangle$ related to
 the constituenr quark formfactor has a simple form $\langle
f_q\rangle\propto\exp(-m_qb/\xi )$.
Following such considerations, the explicit form
 for the generalized
reaction matrix ($U$--matrix) can be constructed and it
 allows
one to obtain the scattering amplitude
 reproducing the main regularities observed in  elastic
scattering at small and large angles \cite{csn}.

The total cross--section has the following energy and quark mass
dependencies \begin{equation} \sigma _{tot}(s)=\frac{\pi  \xi
^2}{\langle m_q \rangle ^2}\Phi (s,N), \label{y} \end{equation} where
$\langle m_q \rangle=\frac{1}{N}\sum_{q=1}^N m_q $ is the mean value
of the constituent quark masses in the colliding hadrons.  The
function $\Phi$ has the following energy dependence:
 \begin{equation} \Phi
(s,N)=\left\{ \begin{array}{cl} \left(8\tilde{g}/N^2\right) \left
[1+N\alpha\sqrt{s}/\langle m_q \rangle \right], & s\ll s_0,\\[2ex]
\ln ^2 s, & s\gg s_0.  \end{array} \right. \label{yy}
\end{equation}

Thus, the $s$--dependence of total cross-section
at $s\ll s_0$ is described by a simple linear function of $\sqrt{s}$.
It has been shown that such dependence is consistent with the
experimental data for the hadron total cross--sections  up to
$\sqrt{s}\sim 0.5$ TeV \cite{preas}.
This is a preasymptotic dependence and it has
nothing to do  with the true asymptotics of the total cross-sections.
In the considered model such behavior of the hadronic cross--sections reflects
 the energy dependence of the number of virtual quarks generated at the
intermediate transient stage of hadronic interaction.

Eqs. (\ref{vdm}), (\ref{vr}) and  (\ref{yy}) provide the following
dependence of $\sigma_{tot}^{\gamma^*p}(W)$:
 \begin{equation}
\sigma_{tot}^{\gamma^*p}(W)=
\left\{ \begin{array}{cl}
 a(Q^2)+b(Q^2)W & W\ll W_0,\\[2ex]
c(Q^2)\ln ^2 W, & W\gg W_0.  \end{array} \right. \label{gg} \end{equation}
where the functions
$a(Q^2)$,
$b(Q^2)$ and
$c(Q^2)$ depend on $Q^2$ and the parameters related to the quark scattering
 in the mean field. $W_0$ separates preasymptotic and asymptotic
region. It was estimated \cite{preas} at the value $W_0\simeq 2$ TeV.
Using the relations between
$\sigma_{tot}^{\gamma^*p}(W)$ and $F_2(x,Q^2)$:
\begin{equation}
\sigma_{tot}^{\gamma^*p}(W)=\frac{4\pi\alpha_{em}}{Q^4}
\frac{4m_p^2x^2+Q^2}{1-x}F_2(x,Q^2)\label{rel}
\end{equation}
and
\[
W^2=Q^2(\frac{1}{x}-1)+m_p^2
\]
it is easily to get the explicit form for $F_2(x,Q^2)$ which is
similar to Eq. (\ref{gg}).  Thus, we can conclude that the $s$--channel
unitarity does provide constraint for $F_2(x,Q^2)$. In the above unitary approach the
preasymptotic behavior \begin{equation}
 F_2(x,Q^2)\propto 1/\sqrt{x} \end{equation}
 at small $x$ is to
be converted into \begin{equation} F_2(x,Q^2)\propto\ln^2( 1/{x})
\end{equation} at
$x\rightarrow 0$.  The line $Q^2=xW_0^2$ separates the asymptotic and
preasymptotic regions.  \section*{Conclusion}

We have demonstrated that under reasonable assumptions the unitarity
restricts the behavior of  the structure function $F_2(x,Q^2)$
at $x\rightarrow 0$. This result is relevant for the region of
not too high $Q^2$ when the fluctuation length is long enough to form bound
state of the two quarks produced by  fluctuating virtual
photon. Usually, direct extension of the on--shell  unitarity
for the virtual photon--hadron scattering by analogy with
hadron--hadron scattering is tacitly implied.  It could not be valid in the
whole kinematical region due to off--shell effects.  We have argued that it
can be justified for the limited region of $Q^2$--values.

 The  treatment of photon--induced reactions similar to the hadron--induced
ones is consistent with the fact that the hadronic as well as (real and
virtual) photon--hadron total cross--sections can be fitted by the same
functional energy dependence in the preasymptotic energy range \cite{preas}.
The original dynamics of hadronic interactions
could be manifested directly
 in the preasymptotic energy region
 while in the region of
very high energies the unitarity plays the most important role.  It screens
the dynamical mechanism and provides a similar behavior for the
cross-sections in the different dynamical approaches.

At high $Q^2$ the photon also can split into $\bar q q$
pair, but this pair will not have enough time to form bound state before
interacting with a proton. In the latter case quarks can not still be
considered as a coherent pair \cite{deld} and due to that the function
$U(s,b)$ is not to be represented as the product of the smeared quark
scattering amplitudes.

  At such high $Q^2$ values the interaction can be treated by
 the perturbative QCD (cf.  e.  g.  \cite{shul}).  Here the off--shell effects
 could play a significant role and
 the  asymptotic behavior of the structure functions at $x\rightarrow 0$ will not
be affected by the screening corrections.
Those corrections will be suppressed according to Eq.  (\ref{**}).

\section*{Acknowledgements} We would like to thank  V. Petrov
for the encouraging discussions.  
\end{document}